\documentclass[11pt]{article}
\pdfoutput=1
\usepackage{jheppub}
\usepackage{amssymb,amsfonts}
\usepackage{mathrsfs}
\usepackage{amsmath}
\usepackage{cleveref}
\let\savenumberline\numberline
\def\numberline#1{\savenumberline{#1.}}
\makeatletter
\renewcommand{\@seccntformat}[1]{\csname the#1\endcsname.\,\,}
\makeatother

\newcommand{\CB}{{\cal B}}

\newcommand{\CG}{{\cal G}}

\newcommand{\rder}{{\overleftarrow{\delta}}}
\renewcommand{\tilde}[1]{\widetilde{#1}}
\renewcommand{\hat}[1]{\widehat{#1}}
\newcommand{\be}{\begin{equation}}
\newcommand{\ee}{\end{equation}}
\newcommand{\bea}{\begin{eqnarray}}
\newcommand{\eea}{\end{eqnarray}}

\newcommand\secref[1]{{\S\ref{#1}}}

\newcommand\tabref[1]{{Table~\ref{#1}}}
\makeatletter
\def\@fpheader{\relax}
\makeatother
\usepackage{graphicx}
\usepackage{latexsym}

\title{\ \vspace{1.5in} \\ Nil-Equivariant Tropological Sigma Models on Filtered Geometries}
\author{Emil Albrychiewicz, Andr\'{e}s Franco Valiente, and Christopher Stites}
\affiliation{\medskip
Leinweber Institute for Theoretical Physics and Department of Physics\\
University of California, Berkeley, CA, 94720-7300, USA\medskip\\
Theoretical Physics Group, Lawrence Berkeley National Laboratory\\
Berkeley, CA 94720-8162, USA}
\emailAdd{ealbrych@berkeley.edu}
\emailAdd{andresfranco@berkeley.edu}
\emailAdd{woeh2owm@berkeley.edu}
\abstract{We investigate the behavior of tropological (tropical topological) sigma models on higher dimensional target spaces and show that higher dimensional spaces explicitly admit nested Maslov dequantizations which lead to nontrivial anisotropic filtration structures. We provide a classification of all inequivalent tropological sigma models that can be constructed for the case of 4D targets and show that, generically, the corresponding sigma-models are not defined on foliated geometries like in the 2D case but instead are defined on filtered manifolds. We find that the nontrivial filtration structures lead to enhanced global symmetries characterized by noncompact nilpotent Lie algebras given by the 4 dimensional step 3 Engel algebra on the space of fields. We provide a Nilmanifold lattice regularization of the noncompact symmetry group and use this Nilmanifold symmetry to construct a natural equivariant extension of the tropological sigma model. We conjecture that these equivariant tropological sigma models are associated with a new version of GW invariants on filtered manifolds known as \textit{filtered Gromov Witten invariants}.
}
\begin{document}
\maketitle
\section{Introduction}
Since their introduction in \cite{gromov, ewtsm}, Gromov-Witten invariants have laid at the heart of modern enumerative geometry and intersection theory. In mathematics, they are often introduced as intersection numbers on the moduli space of pseudoholomorphic maps from a Riemann surface $\Sigma_{g, n}$, with genus $g$ and $n$ punctures, into a target space $X$. In quantum field theory, it has been shown that they physically emerge as the exact non-perturbative correlators of type A topological sigma models through localization methods and have been shown to be a useful tool for testing dualities \cite{Aganagic:2003db} using mirror symmetry \cite{Kontsevich:1994dn, ewmirror}, calculating worldsheet instanton corrections and encoding BPS spectra \cite{Frenkel:2006fy, Frenkel:2008vz}. 

In \cite{trsm}, it was pointed out that Mikhalkin's result \cite{mikhalkin} which states that the Gromov-Witten invariants may be calculated by investigating the tropical limit of the underlying geometries \cite{msintro, rau, mikhalkinrau} was carried out in the path integral perspective. From a mathematician's perspective, the tropicalization of a complex variety essentially reduces down the geometry into a piece-wise linear real algebraic geometry known as a tropical geometry by effectively forgetting the phase of the complex coordinates parameterizing the complex variety; this allows one to replace the difficult analysis of psuedoholomorphic curves with simpler combinatorial questions of the underlying tropical geometry. In order to implement this limit from the path integral perspective, it becomes essential to understand how to construct a limit that encodes tropical geometry without losing standard path integral techniques by working in the tropical semiring. For the case of the standard topological sigma A model, the tropical limit converges to an anisotropic Witten-type topological field theory described by worldsheet foliations that was coined a \textit{tropological} (tropical topological)  sigma model.  

In further developments, it has been shown \cite{Albrychiewicz:2025hzt} that these tropological sigma model admit both tropical Lagrangian branes and tropical coisotropic branes constructions. Despite these results, the explicit constructions have only been carried out for low-dimensional target space geometries such as $\mathbb{CP}^1$ and the cotangent bundle of the 2-torus $T^*T^2$. The tropical limit of these target spaces were shown to be geometries that are characterized by nilpotent endomorphisms of the tangent bundle known as Jordan structures which induce natural foliations.

In this paper, we study tropological sigma models on the manifold $\mathbb{CP}^2$ and investigate the many inequivalent ``tropical limits" that one may consider. We demonstrate that a careful choice of tropical limit results in a target space geometry that is not simply a foliated manifold but instead it is a filtered geometry equipped with a non-holonomic tangent space induced by the filtration. This filtration structure can be seen as an additional global symmetry that acts on the space of fields. For the case of $\mathbb{CP}^2$, we find that this additional symmetry algebra can be identified as a noncompact finite-dimensional nilpotent step 3 Lie algebra known as an Engel algebra. The results are suggestive that tropological sigma models might be able to extract additional invariants beyond the Gromov-Witten invariants for tropicalized Jordan structures.

Tropological sigma models were introduced in \cite{trsm}, where their BRST/path-integral construction and a range of low-dimensional examples were developed. The present paper extends that framework to the case of $\mathbb{CP}^2$ and to higher-dimensional Jordan structures. It develops a four-real-dimensional example coming from a complex target space, shows that $\mathbb{CP}^2$ admits inequivalent tropical limits leading to genuinely inequivalent tropological sigma models rather than alternative presentations of the same theory, and exhibits a nested two-step Maslov dequantization producing the $J_{3,1}$ sector. This sector is shown to be naturally associated with filtered, rather than merely foliated, geometry, together with an Engel-type nilpotent symmetry on the space of fields. The paper also regularizes this noncompact symmetry by passing to a nilmanifold quotient, thereby obtaining a nil-equivariant extension of the tropological sigma model. In this setting, the discussion of filtered Gromov--Witten invariants in \S4 is most naturally understood as a conjectural consequence of the new equivariant structure introduced here, rather than as part of the framework already established in \cite{trsm}.

In \secref{sec:TropSigma}, we provide a quick systematic review of how the tropical limit is implemented in the path integral and show that in low dimensions, the canonical geometric data that tropical field theories are associated with are nilpotent endomorphisms of the tangent bundle known as Jordan structures. We present a classification of Jordan structures in higher dimensions based on the Jordan block decomposition. We argue that the Jordan structure of interest $J_{3,1}$ in this paper can be arrived at through a double Maslov dequantization limit which encodes nested foliations that result in a nontrivial filtration structure. In \secref{sec:Actions}, we will give the explicit localization equations and actions for each of Jordan structures. We find that the tropological sigma model associated to $J_{3,1}$ gives rise to the noncompact nilpotent Lie algebra mentioned previously. In \secref{sec:NilTrop}, we show how to regularize the noncompactness of the symmetry group by taking a lattice quotient regularization which yields a nilmanifold global symmetry and construct the equivariant extension of the tropological sigma model. We construct equivariant BRST cohomology and observables. We conclude in \secref{sec:Conclusions} with a discussion of what new features should be expected in higher dimensions beyond 4 as well as natural follow up questions associated to the Nil-equivariant tropological sigma model.

\section{Tropological Sigma Models and Higher Dimensional Jordan Structures}
\label{sec:TropSigma}

For a detailed introduction to tropological sigma models, we refer readers to \cite{trsm}. Here, we limit review to the standard constructions and notations needed to construct tropological sigma models for 2-dimensional target spaces and then discuss how to naturally extend this to higher dimensional target spaces.

\subsection{A Review on Tropological Sigma Models}
The practical implementation of the tropical limit is done locally through the Maslov dequantization \cite{msintro, litvinov, viro, virohyper} where one begins by equipping a complex manifold (or complex algebraic variety) $\hat\Sigma$ with local complex coordinates $z_i$ and deforming the local coordinates as
\begin{equation}
z=e^{\frac{r}{\hbar}+i \theta},
\end{equation}
where for now we limit the discussion to one complex dimension. We then say the tropical limit is taken as $\hbar\rightarrow 0$. Generically, we use hats on geometric objects prior to the tropical limit and remove the hat once the tropical limit has been taken. We call the $(r,\theta)$, the adapted coordinates. Tropicalization is generically approached by logarithmically scaling the modulus $|z|$ resulting in a real algebraic variety whose underlying operations are given by the tropical semiring $\mathbb{T}$ where the addition is replaced by a maximum (or minimum depending on the convention) operation and the multiplication becomes the standard addition
\begin{equation}
\begin{aligned}
& x \oplus y=\operatorname{max}(x, y), \\
& x \otimes y=x+y.
\end{aligned}
\end{equation}
We call this approach where the phase of $z$ is ``forgotten", the quotient space approach. In this paper, the Maslov dequantization allows us to access the \textit{covering space approach} where we keep the additional angular variable and retain the differential structure of a complex manifold while still encoding the underlying tropical properties. The upshot of this approach is that we are still able to utilize all standardized path integral methods for the purposes of practical calculations, in particular, we can still unambiguously write down derivatives.

Using the Maslov dequantization, one can show that the tropological sigma model may be obtained as a non-relativistic limit of the relativistic sigma model directly through a limit of the action or indirectly by first taking a tropical limit of the localization equations and reconstructing a Witten-type topological field theory through standard BRST cohomological arguments. The tropical limit is done by performing a Maslov dequantization on both the worldsheet $\hat{\Sigma}$ and the target space $\hat{X}$ coordinates.

The topological A-model comes equipped with a worldsheet complex structure $\hat\epsilon$ on its worldsheet $\hat\Sigma$ and is often studied on target spaces $\hat{X}$ which are complex manifolds with an almost complex structure denoted $\hat{J}$. Taking the tropical limit of these objects in the holomorphic basis lead to a nilpotent endomorphism of the tangent bundle known as a Jordan structure $\epsilon: T \Sigma \rightarrow T \Sigma$, whose action in the adapted coordinates $(r,\theta)$ are given by
\begin{equation}
\epsilon\left(\partial_r\right)=\partial_\theta, \quad \epsilon\left(\partial_\theta\right)=0.
\end{equation}
One can straightforwardly check that this implies $\epsilon^2=0.$ We want to emphasize that unlike the standard complex structure which gives a Dolbeault decomposition of the tangent bundle into the holomorphic and anti-holomorphic tangent bundle, the Jordan decomposition instead gives us a two-step filtration on the tangent space of the worldsheet
\begin{equation}
0 \subset \operatorname{ker}(\epsilon)=\operatorname{im}(\epsilon) \subset T \Sigma.
\end{equation}
For the case of two-dimensional worldsheets, this filtration induces a natural foliation. A natural consequence of this new foliation is that the usual diffeomorphism symmetry of the A-model gets deformed into foliation-preserving diffeomorphisms which preserve the Jordan structure
\begin{equation}
\begin{aligned}
& \widetilde{r}=\widetilde{r}(r), \\
& \widetilde{\theta}=\widetilde{\theta}_0(r)+\theta \partial_r \widetilde{r}(r) .
\end{aligned}
\end{equation}
If we denote the local complex coordinates on the target space by $Z$, then we will denote the adapted coordinates of the Maslov dequantization on the target space by $\vec{Y}=(X,\Theta)$ i.e.,
\begin{equation}
Z=\exp \left\{\frac{X}{\hbar}+i \Theta\right\}.
\end{equation}
The localization equations of the tropological sigma model can then be obtained by directly taking a tropical limit of the pseudoholomorphicity equations of the topological sigma model resulting in
\begin{equation}
{E}_\alpha{ }^i={\varepsilon}_\alpha{ }^\beta \partial_\beta Y^i-{J}_j{ }^i \partial_\alpha Y^j=0,
\end{equation}
more explicitly, the system of differential equations decomposes into
\begin{equation}
E_r^X=\partial_\theta X=0, \quad E_r^{\Theta}=\partial_\theta \Theta-\partial_r X=0, \quad E_\theta^X=0, \quad E_\theta^{\Theta}=-\partial_\theta X=0.
\end{equation}
A short analysis of these equations reveal that there is an additional target space symmetry known as the $
\alpha$ symmetry which can be parametrized by two projectable functions $\alpha_0(r),\alpha_1(r)$ given by
\begin{equation}
\begin{aligned}
\delta X & =\alpha_1(r), \\
\delta \Theta & =\alpha_0(r)+\theta \partial_r \alpha_1(r).
\end{aligned}
\end{equation}
Once all the geometric ingredients have been laid out, the action can now be constructed through constructing a BRST differential $Q$ and an appropriate gauge fixing fermion. The antighost field $\chi^\alpha{ }_i$ is related to an Grassmann even auxiliary tensor density field $\mathcal{B}^{\alpha}_i$ and satisfy
\begin{equation}
\begin{aligned}
\left\{Q, \chi^\alpha{ }_i\right\} & =\mathcal{B}^\alpha{ }_i,\\
{\left[Q, \mathcal{B}^\alpha{ }_i\right] } & =0,
\end{aligned}
\end{equation}
a BRST representative for the action for the tropological sigma model may then be summarized as
\begin{equation}
\label{eqn:TropAction}
\mathcal{S}=\int_{\Sigma} d^2 \sigma \mathcal{B}_{\;i}^\alpha E_{\;\alpha}^i.
\end{equation}
In the next section, we will employ precisely this construction in order to construct tropological sigma models on higher dimensional target spaces. For completeness, we mention that a careful analysis of the observables then shows that the tropological sigma model correctly reproduces the Gromov-Witten invariants and demonstrates \cite{sequel} a straightforward coupling to non-relativistic topological gravity whose dynamical variables are vielbeins that encode dynamical foliations. In addition, these tropological sigma models admit topological brane constructions which are analogous to the Lagrangian and coisotropic branes as well as tropical (non-topological) branes from which result from a boundary conformal algebra related to the analytic continuation of the tropological sigma model \cite{Albrychiewicz:2024tqe}. These boundary algebra are reminiscent to the same Carollian algebras that have connections to flat space holography \cite{Susskind:1998vk} and tensionless string theories \cite{Gross:1987kza, Gross:1987ar, ziqi}.

\subsection{Higher Order Jordan Structures}
The tropical limit is implemented locally by Maslov dequantization, with the resulting geometric datum given by a nilpotent endomorphism of the tangent bundle, i.e. a Jordan structure. For the $4$-real-dimensional case relevant to $\mathbb{CP}^2$, the natural starting point is therefore a classification of the possible nilpotent degenerations of a complex structure, first in two real dimensions and then in four. This classification supplies the geometric framework used in the next section to construct the explicit $\mathbb{CP}^2$ examples, where different Jordan types give rise to inequivalent tropological sigma models.

In two real dimensions, there is a unique complex structure up to a change of basis that you can equip $\mathbb{C}$ with, a matrix representation for the complex structure may be given by
\begin{equation}
\left(\begin{array}{ll}
0 & 1 \\
-1 & 0
\end{array}\right).
\end{equation}
Taking the tropical limit gives a nilpotent endomorphism of nilpotency of a degree two. We will call this a \textit{second order Jordan structure} where the second order refers to the nilpotency degree. Standard arguments in a finite-dimensional linear algebra shows that every nontrivial nilpotent endomorphism of degree 2 is conjugate to a Jordan block 
\begin{equation}
J_{2}=\left(\begin{array}{ll}
0 & 1 \\
0 & 0
\end{array}\right),
\end{equation}
which is unique, again up to a change of basis. The subscript $2$ on $J$ refers to the size of the Jordan block describing the Jordan structure. The most natural way to classify Jordan structures on $\mathbb{C}^n$ for $n\geq 2$ is to use the Jordan block classification of nilpotent matrices.  

For the case of $\mathbb{C}^2$, we can give an explicit matrix representation of second order Jordan structures in terms of 4x4 matrices. The only possible nonzero Jordan block decompositions are given by either one block of size two along with two blocks of size 1 or two blocks of size 2. The first possibility has a matrix representation of the form
\begin{equation}
\label{eqn:J211}
J_{2,1,1}=\left(\begin{array}{llll}
0 & 1 & 0 & 0 \\
0 & 0 & 0 & 0 \\
0 & 0 & 0 & 0 \\
0 & 0 & 0 & 0
\end{array}\right).
\end{equation}
We will use the notation $J_{a,b,c}$ where the indices correspond to the size of the Jordan block and the number of Jordan blocks is given by the number of separated indices. The linear map corresponding to this matrix has rank 1 (in general Jordan block of order $k$ has rank $k-1$) and a 3-dimensional kernel. For the case of two blocks of size 2, we have
\begin{equation}
\label{eqn:J22}
    J_{2,2}=\left(\begin{array}{llll}
0 & 1 & 0 & 0 \\
0 & 0 & 0 & 0 \\
0 & 0 & 0 & 1 \\
0 & 0 & 0 & 0
\end{array}\right),
\end{equation}
here the map has rank 2 and a 2-dimensional kernel. These two Jordan structures are not conjugate to each other, and they exhaust all possibilities for a nilpotent endomorphism of degree 2 in 4 dimensions. 

However, it turns out that there is another possibility that was mentioned in \cite{trsm} and we can have Jordan structures of nilpotency degree 3. In other words, we can consider \textit{third order} Jordan structures. In a 4-dimensional vector space, a nilpotent endomorphism of degree 3 is one for which $J^3=0$ but $J^2 \neq 0$. In the Jordan classification, the nilpotency index is given by the size of the largest Jordan block. Since the overall dimension is 4, the only possible Jordan block decomposition that achieves this is one Jordan block of size 3 and one block of size 1. Up to conjugation, this form is unique. Thus, there is exactly 1  nilpotent endomorphism of degree 3 on a 4 -dimensional vector space. A matrix representation can be given by
\begin{equation}
\label{eqn:J31}
J_{3,1}=\left(\begin{array}{llll}
0 & 1 & 0 & 0 \\
0 & 0 & 1 & 0 \\
0 & 0 & 0 & 0 \\
0 & 0 & 0 & 0
\end{array}\right).
\end{equation}

Lastly, we can extend this up to the case of a fourth order Jordan structure. This condition forces the largest Jordan block of $J$ to have size 4. In a 4-dimensional space, the only possibility is to have one single Jordan block of size 4. Thus, up to conjugation, there is exactly one such nilpotent endomorphism. A canonical matrix representation with respect to a suitable basis is:
\begin{equation}
\label{eqn:J4}
J_4=\left(\begin{array}{llll}
0 & 1 & 0 & 0 \\
0 & 0 & 1 & 0 \\
0 & 0 & 0 & 1 \\
0 & 0 & 0 & 0
\end{array}\right).
\end{equation}

We cannot obtain any higher order Jordan structures for a vector space of dimension 4 but the construction readily generalizes to higher dimensions. In principle, we are able to equip the target space with any of these Jordan structures and study the tropological sigma model on these target spaces but a natural question is: are any of these Jordan structures singled out and which is the natural one to to equip $\mathbb{CP}^2$ with?

It turns out that there is exactly one inequivalent complex structure on a 2-complex dimensional vector space up to a complex linear equivalence. Choosing a matrix representation gives
\begin{equation}
\hat{J}=\left(\begin{array}{cccc}
0 & 1 & 0 & 0 \\
-1 & 0 & 0 & 0 \\
0 & 0 & 0 & 1 \\
0 & 0 & -1 & 0
\end{array}\right).
\end{equation}
Notice that this higher dimensional complex structure is essentially the tensor product of two lower dimensional complex structures. We will see that we can still construct a unique tropicalization in 4 dimensions. We denote the degenerations by primes. If you take the tropical limit of the first embedded complex structure and leave the second embedded complex structure alone,  we obtain 
\begin{equation}
J'=\left(\begin{array}{cccc}
0 & 1 & 0 & 0 \\
0 & 0 & 0 & 0 \\
0 & 0 & 0 & 1 \\
0 & 0 & -1 & 0
\end{array}\right),
\end{equation}
which is not nilpotent and hence is not a Jordan structure. Taking the tropical limit of the 2nd complex structure and leaving the first one alone is equivalent up to a matrix transformation to the first option. One obtains
\begin{equation}
J''=\left(\begin{array}{cccc}
0 & 1 & 0 & 0 \\
-1 & 0 & 0 & 0 \\
0 & 0 & 0 & 1 \\
0 & 0 & 0 & 0
\end{array}\right)
\end{equation}
One can easily find a similarity transform which tells us that $J'$ is equivalent to $J''$ up to a linear change of coordinates.

Taking a tropical limit such that both embedded complex structures become embedded Jordan structures, we obtain
\begin{equation}
J'''=\left(\begin{array}{cccc}
0 & 1 & 0 & 0 \\
0 & 0 & 0 & 0 \\
0 & 0 & 0 & 1 \\
0 & 0 & 0 & 0
\end{array}\right).
\end{equation}
This linear map is nilpotent and represents a true Jordan structure, hence we see that $J'''=J_{2,2}$  is uniquely picked out from tropicalizing the complex structure on $\mathbb{CP}^2$. We examine all of these Jordan structures in the next section, despite $J_{2,2}$ being picked out from a direct tropical limit. It will turn out that $J_{3,1}$ is the most interesting Jordan structure. The reason for this is due to the fact that one cannot directly arrive at $J_{3,1}$ through a single direct Maslov dequantization. Instead, one must carefully perform two distinct Maslov dequantizations in series in such a way that that two of the entries of the complex structure is killed off and we have $J^3=0$. One can achieve this by first performing a Maslov dequantization to zero out one of the entries and obtain a matrix equivalent to $J_4$, rotating the coordinates and then doing another Maslov dequantization limit kills another entry which results in $J_{3,1}$. Thus, in general, we would expect that each Maslov dequantization induces a distinct foliation structure that when overlapping manifests itself as a nontrivial filtration on the space. Such a Jordan structure has first been described in the 3-dimensional target space case in \cite{trsm}.  We discuss this nontrivial filtration structure in \secref{sec:NonTrivialFerm}. 

\section{Actions for Tropological Sigma Models on $\mathbb{CP}^2$}
\label{sec:Actions}

We now turn to the explicit tropological sigma models associated with the relevant $4$-dimensional Jordan structures in the case of $\mathbb{CP}^2$. Throughout this section the target space is $\hat X$, which has $4$ real dimensions, with local complex coordinates $Z^i,\bar Z^i$, $i=1,2$. We use adapted worldsheet coordinates $(r,\theta)$ and adapted target-space coordinates $Y^i=(X^i,\Theta^i)$, and implement the Maslov dequantization by
\[
Z^i = e^{X^i/\hbar + i\Theta^i},
\]
followed by the limit $\hbar\to0$. For $\mathbb{CP}^2$, the Jordan sectors $J_{2,2}$, $J_4$, and $J_{3,1}$ already lead to inequivalent localization equations and BRST-type actions. The trivial case $J_{2,1,1}$ in \eqref{eqn:J211}, which reproduces the original tropological sigma model on tropical projective space $\mathbb{TP}^1$, will therefore be omitted, and attention will be restricted to the three nontrivial cases $J_{2,2}$, $J_4$, and $J_{3,1}$ in \eqref{eqn:J22}--\eqref{eqn:J4}.

The action of the nilpotent BRST charge on the space of fields is given by
\begin{equation}
\label{eq:BRST_action}
\begin{alignedat}{2}
[Q,X^1]       &= \psi^1, \quad & [Q,X^2]       &= \psi^2,\\
[Q,\Theta^1] &= \varphi^1, \quad & [Q,\Theta^2] &= \varphi^2.
\end{alignedat}
\end{equation}
additionally, their action on the foliation-preserving diffeomorphism ghosts $(\psi^1,\psi^2,\varphi^1,\varphi^2)$ is given by
\begin{equation}
\left\{Q, \varphi^1\right\}=\left\{Q, \varphi^2\right\}=\left\{Q, \psi^1\right\}=\left\{Q, \psi^2\right\}=0.
\end{equation}
The bosonic auxiliary fields $\mathcal{B}^\alpha_i$ and its antighost superpartner $\chi^\alpha_i$ are defined as usual and put into a BRST multiplet
\begin{equation}
\begin{aligned}
& \left\{Q, \chi_{\;I}^\alpha\right\}=\mathcal{B}_{\;I}^\alpha, \\
& \left\{Q, \mathcal{B}_{\;I}^\alpha\right\}=0,
\end{aligned}
\end{equation}
and the action is defined as in \eqref{eqn:TropAction}. This action is invariant under an additional gauge symmetry that preserves $Y^i$ but modifies $\CB^{\alpha}_{\;i}$. We can fix this gauge symmetry in a such way that we preserve conformal invariance of the worldsheet, however, the full conformal invariance of the target space cannot be preserved. We refer reader for further details to \cite{trsm}, and in what follows we pick guage fixing such that
\begin{align}
    \label{eqn:GaugeB}
    \mathcal{B}_{\;X^1}^\alpha=\mathcal{B}_{\;X^2}^\alpha & =0.
\end{align}
The antighosts $\chi^{\alpha}_{\; i}$, which follow the same pattern as of the auxiliaries, are similarly gauge fixed with the choice
\begin{align}
    \label{eqn:GaugeChi}
    \chi_{\;X^1}^\alpha=\chi_{\;X^2}^\alpha & =0.
\end{align}
The leftover non-zero components of these tensor densities in adapted coordinates are denoted as
\begin{equation}
\begin{aligned}
\label{eqn:NonZeroComp}
\mathcal{B}_{\;\Theta^1}^r & =B_1, \quad \mathcal{B}_{\;\Theta^2}^r =B_2, \quad
\mathcal{B}_{\;\Theta^1}^\theta =-\beta_1, \quad
\mathcal{B}_{\;\Theta^2}^\theta =-\beta_2, \\
\chi_{\;\Theta^1}^r & =\Xi_1, \quad
\chi_{\;\Theta^2}^r  =\Xi_2, \quad 
\chi_{\;\Theta^1}^\theta  =-\chi_1, \quad 
\chi_{\;\Theta^2}^\theta  =-\chi_2.
\end{aligned}
\end{equation}
To integrate out half of auxiliaries, we can add a BRST invariant and conformally invariant term of the form
\begin{align}
    -\frac{1}{2}\int d^2\sigma\gamma_{\alpha\beta}\left(H^{\Theta^1\Theta^1}\CB^{\alpha}_{\;\Theta^1}\CB^{\beta}_{\;\Theta^1}+H^{\Theta^2\Theta^2}\CB^{\alpha}_{\;\Theta^2}\CB^{\beta}_{\;\Theta^2}\right)=-\frac{1}{2}\int d^2\sigma\left(B_1^2+B_2^2\right),
\end{align}
where $\gamma_{\alpha\beta}$ is a tensor density of weight $-1$ constructed with a worldsheet metric and in the adapted coordinates its leftover component after tropicalization is $\gamma_{rr}$. $H^{\Theta^1\Theta^1}$ and $H^{\Theta^2\Theta^2}$ are the non-zero components of the target space metric.
However, we cannot add to the action a similar quadratic term for $\beta_1$ and $\beta_2$ as such term would break worldsheet conformal invariance.

\subsection{Tropological Sigma Model on $\mathbb{TP}^2 $ with $J_{2,2}$ }
As argued in the previous section, the Jordan structure that is singled from the tropical limit of $\mathbb{CP}^2$ is
\begin{equation}
J_{2,2}=\left(\begin{array}{llll}
0 & 1 & 0 & 0 \\
0 & 0 & 0 & 0 \\
0 & 0 & 0 & 1 \\
0 & 0 & 0 & 0
\end{array}\right),
\end{equation}
we view tropical projective space $\mathbb{TP}^2$ as having the same underlying differential structure as $\mathbb{CP}^2$ but instead of being equipped with its standard complex structure, we equip it with $J_{2,2}$. The localization equations are then

\begin{equation}
\label{eq:localization_4x2}
\begin{alignedat}{2}
E_{\;\;r}^{X^1}      &= \partial_\theta X^1,                     &\quad
E_{\;\;\theta}^{X^1}      &= 0,\\
E_{\;\;r}^{\Theta^1} &= \partial_\theta \Theta^1 - \partial_r X^1, &\quad
E_{\;\;\theta}^{\Theta^1} &= -\,\partial_\theta X^1,\\
E_{\;\;r}^{X^2}      &= \partial_\theta X^2,                     &\quad
E_{\;\;\theta}^{X^2}      &= 0,\\
E_{\;\;r}^{\Theta^2} &= \partial_\theta \Theta^2 - \partial_r X^2, &\quad
E_{\;\;\theta}^{\Theta^2} &= -\,\partial_\theta X^2,
\end{alignedat}
\end{equation}
we find that our corresponding action is 
\begin{equation*}
\begin{aligned}
S & =\frac{1}{e} \int d r d \theta\left\{Q, \Xi_j\left(\partial_\theta \Theta^j-\partial_r X^j-\frac{1}{2} B^j\right)+\chi_j \partial_\theta X^j\right\} \\
& =\frac{1}{e} \int d r d \theta\left[B_j\left(\partial_\theta \Theta^j-\partial_r X^j\right)+\beta_j \partial_\theta X^j-\Xi_j\left(\partial_\theta \varphi^j-\partial_r \psi^j\right)-\chi_j \partial_\theta \psi^j-\frac{1}{2} B_j B^j\right]
\end{aligned}
\end{equation*}
Here $e$ is a coupling constant of this quantum theory. This action is quadratic in the auxiliary fields $B_j$ and integrating them out results in the final action
\begin{align*}
S&=\frac{1}{e} \int d r d \theta\Big[\frac{1}{2}\left(\partial_\theta \Theta^1-\partial_r X^1\right)^2+\frac{1}{2}\left(\partial_\theta \Theta^2-\partial_r X^2\right)^2+\beta_j \partial_\theta X^j \\
&-\Xi_j\left(\partial_\theta \varphi^j-\partial_r \psi^j\right)-\chi_j \partial_\theta \psi^j\Big]
\end{align*}
This action is simply a doubling of the standard tropological sigma model on $\mathbb{TP}^1$ that is introduced in \cite{trsm}. It's clear from the form of the Jordan structure that the underlying geometry is that of a foliated manifold where we have individual foliations running over the $(X_1,\Theta^1)$ and $(X_1,\Theta^2)$ coordinates. Given this doubling, it's clear from previous arguments that the correlation functions of this theory will match with the standard Gromov-Witten invariants of $\mathbb{CP}^2$.

\subsection{Tropological Sigma Model on $\mathbb{TP}^2 $ with $J_{4}$  }
Now, we consider the fourth order Jordan structure
\begin{equation}
J_4=\left(\begin{array}{llll}
0 & 1 & 0 & 0 \\
0 & 0 & 1 & 0 \\
0 & 0 & 0 & 1 \\
0 & 0 & 0 & 0
\end{array}\right).
\end{equation}
The corresponding localization equations are
\begin{equation}
\label{eq:localization_4x2_second}
\begin{alignedat}{2}
E_{\;\;r}^{X^1}      &= \partial_\theta X^1,                          &\quad
E_{\;\;\theta}^{X^1}      &= 0,\\
E_{\;\;r}^{\Theta^1} &= \partial_\theta \Theta^1 - \partial_r X^1,   &\quad
E_{\;\;\theta}^{\Theta^1} &= -\,\partial_\theta X^1,\\
E_{\;\;r}^{X^2}      &= \partial_\theta X^2 - \partial_r \Theta^1, &\quad
E_{\;\;\theta}^{X^2}      &= -\,\partial_\theta \Theta^1,\\
E_{\;\;r}^{\Theta^2} &= \partial_\theta \Theta^2 - \partial_r X^2,   &\quad
E_{\;\;\theta}^{\Theta^2} &= -\,\partial_\theta X^2,
\end{alignedat}
\end{equation}
and the action is
\begin{equation*}
S=\frac{1}{e} \int d r d \theta\left[\frac{1}{2}\left(\partial_\theta \Theta^1-\partial_r X^1\right)^2+\frac{1}{2}\left(\partial_\theta \Theta^2-\partial_r X^2\right)^2+\beta_j \partial_\theta X^j-\Xi_j\left(\partial_\theta \varphi^j-\partial_r \psi^j\right)\right].
\end{equation*}
The localization equations may be solved locally on a worldsheet patch and yield the solutions
\begin{equation}
\label{eq:split_X_Theta}
\begin{alignedat}{2}
X^1        &= f_1(r),                \quad & 
X^2        &= f_3(r), \\[6pt]
\Theta^1   &= f_2(r),                \quad &
\Theta^2   &= f_4(r) + \theta\,\partial_r f_3(r),
\end{alignedat}
\end{equation}
where the $f_j(r)$ are arbitrary projectable functions. Interestingly, unlike the tropological sigma model for the case where $J_{2,2}$ where we got a doubling, the structure of $J_4$ seems to totally collapse the $(X^1,\Theta^1)$ directions and only leaves room for a tropical winding number for the $\Theta^2$ direction. If we assume that the target space $\Theta^2$ is periodic, then we obtain a winding number $w$ given by
\begin{equation}
\begin{aligned}
& X^2(r)=wr \\
& \Theta^2(r,\theta)=f_4(r)+w \theta ,
\end{aligned}
\end{equation}
It's clear from the local solutions that if the $\Theta^1$ coordinate is also periodic, we won't get a doubling of winding numbers.

\subsection{Tropological Sigma Model on $\mathbb{TP}^2$ with $J_{3,1}$: Nilpotent Fermionic Symmetries}
\label{sec:NonTrivialFerm}
The localization equations associated to
\begin{equation}
J_{3,1}=\left(\begin{array}{llll}
0 & 1 & 0 & 0 \\
0 & 0 & 1 & 0 \\
0 & 0 & 0 & 0 \\
0 & 0 & 0 & 0
\end{array}\right),
\end{equation}
are
\begin{equation}
\label{eq:localization_4x2_final}
\begin{alignedat}{2}
E_{\;\;r}^{X^1}      &= \partial_\theta X^1,                       &\quad
E_{\;\;\theta}^{X^1}      &= 0,\\
E_{\;\;r}^{\Theta^1} &= \partial_\theta \Theta^1 \;-\;\partial_r X^1, &\quad
E_{\;\;\theta}^{\Theta^1} &= -\,\partial_\theta X^1,\\
E_{\;\;r}^{X^2}      &= \partial_\theta X^2 \;-\;\partial_r \Theta^1,&\quad
E_{\;\;\theta}^{X^2}      &= -\,\partial_\theta \Theta^1,\\
E_{\;\;r}^{\Theta^2} &= \partial_\theta \Theta^2,                   &\quad
E_{\;\;\theta}^{\Theta^2} &= 0,
\end{alignedat}
\end{equation}
and the action
\begin{equation}
\begin{aligned}
\label{eqn:ActionJ31}
S & =\frac{1}{e} \int_{\Sigma} d r d \theta\left\{Q, \chi_{\;\;i}^\alpha E_{\;\alpha}^i-\frac{1}{2}\Xi_i B^i\right\} \\
& =\frac{1}{e} \int_{\Sigma} d r d \theta\Big[B_1\left(\partial_\theta \Theta^1-\partial_r X^1\right)+B_2\left(\partial_\theta \Theta^2\right)+\beta_1\left(\partial_\theta X^1\right) \\
&-\frac{1}{2} (B_1^2+B_2^2)-\Xi_1\left(\partial_\theta \varphi^1-\partial_r \psi^1\right)- \Xi_2\left(\partial_\theta \varphi^2\right)-\chi_1\left(\partial_\theta \psi^1\right)\Big].
\end{aligned}
\end{equation}
Integrating out the $B_1$ and $B_2$ auxiliary fields, one can obtain the following action
\begin{equation}
\begin{aligned}
S & =\frac{1}{e} \int_{\Sigma} d r d \theta\left[\frac{1}{2}\left(\partial_\theta \Theta^1-\partial_r X^1\right)^2+\frac{1}{2}\left(\partial_\theta \Theta^2\right)^2+\beta_1\left(\partial_\theta X^1\right)-\Xi_1\left(\partial_\theta \varphi^1-\partial_r \psi^1\right)-\right. \\
& \left.\Xi_2\left(\partial_\theta \varphi^2\right)-\chi_1\left(\partial_\theta \psi^1\right)\right].
\end{aligned}
\end{equation}
Unlike the previous tropological sigma models, this one has an additional infinitesimal symmetry which generalizes the $\alpha$ symmetry. The symmetry generators will have a trivial action on some fields but a nontrivial action on others, we will list out only the nontrivial transformations. The first symmetry generator denoted $\delta_1$ gives 
\begin{equation}
\begin{aligned}
\delta_1 X^1 & =\psi^1, \\
\delta_1 \Theta^1 & =\varphi^1, \\
\delta_1 \Xi_1 & =\partial_\theta \Theta^1-\partial_r X^1, \\
\delta_1 \chi_1 & =\beta_1.
\end{aligned}
\end{equation}
The second symmetry generator is given by
\begin{equation}
\begin{aligned}
& \delta_2 \varphi^1=-\varphi^2, \\
& \delta_2 \Xi_2=\Xi_1.
\end{aligned}
\end{equation}
In addition to $\delta_1$ and $\delta_2$, one is able to generate an independent symmetry generator by taking the appropriate graded commutators $\left[\delta_1, \delta_2\right]:=\delta_3$, which gives
\begin{equation}
\begin{aligned}
& \delta_3 \Theta^1=\varphi^2, \\
& \delta_3 \Xi_2=\partial_\theta \Theta^1-\partial_r X^1,
\end{aligned}
\end{equation}
as well as $\left[\delta_1, \delta_3\right]:=\delta_4$ which gives
\begin{equation}
\begin{aligned}
& \delta_4 \Xi_1=-\partial_\theta \varphi^2, \\
& \delta_4 \Xi_2=\partial_\theta \varphi^1-\partial_r \psi^1.
\end{aligned}
\end{equation}
We can more explicitly write down the generators $Q_i$ by using functional derivatives acting to the left over the foliated worldsheet
\begin{equation}
Q_i=\int d^2 \sigma \delta_i \Phi^A(\sigma) \frac{\rder}{\delta \Phi^A(\sigma)},
\end{equation}
where $\Phi^A$ denotes any field and we included the arrow to underline that the derivative acts from the right. Then, the fermionic symmetry generators may be written down as
\begin{equation}
\begin{gathered}
Q_1=\int d^2 \sigma\left[\psi^1 \frac{\rder}{\delta X^1}+\varphi^1 \frac{\rder}{\delta \Theta^1}+\left(\partial_\theta \Theta^1-\partial_r X^1\right) \frac{\rder}{\delta \Xi_1}+\beta_1 \frac{\rder}{\delta \chi_1}\right], \\
Q_2=\int d^2 \sigma\left[-\varphi^2 \frac{\rder}{\delta \varphi_1}+\Xi_1 \frac{\rder}{\delta \Xi_2}\right], \\
Q_3=\int d^2 \sigma\left[\varphi^2 \frac{\rder}{\delta \Theta^1}+\left(\partial_\theta \Theta^1-\partial_r X^1\right) \frac{\rder}{\delta \Xi_2}\right], \\
Q_4=\int d^2 \sigma\left[-\partial_\theta \varphi^2 \frac{\rder}{\delta \Xi_1}+\left(\partial_\theta \varphi^1-\partial_r \psi^1\right) \frac{\rder}{\delta \Xi_2}\right].
\end{gathered}
\end{equation}
One can show that these differential operators generate the infinitesimal symmetries and they satisfy 
\begin{align}
\label{eqn:ChargeComm}
    \left[Q_1, Q_2\right]=Q_3, \quad \left[Q_1, Q_3\right]=Q_4,
\end{align}
with all other commutators vanishing. The symmetry also holds off-shell, if we consider \eqref{eqn:ActionJ31}, where auxiliary fields have not been integrated out yet, we find a simple modification of charges
\begin{align*}
& Q_1'=\int d^2 \sigma\left[\psi^1 \frac{\rder}{\delta X^1}+\varphi^1 \frac{\rder}{\delta \Theta^1}+(\partial_\theta \Theta^1-\partial_r X^1) \frac{\rder}{\delta \Xi_1}+\beta_1 \frac{\rder}{\delta \chi_1}+\left(\partial_\theta \varphi^1-\partial_r \psi^1\right) \frac{\rder}{\delta B_1}\right], \\
& Q_3'=\int d^2 \sigma\left[\varphi^2 \frac{\rder}{\delta \Theta^1}+ (\partial_\theta \Theta^1-\partial_r X^1)\frac{\rder}{\delta \Xi_2}+\partial_\theta \varphi^2\frac{\rder}{\delta B_1}\right],
\end{align*}
with $Q_2$ and $Q_4$ not modified. These charges again satisfy \eqref{eqn:ChargeComm}.

The symmetry algebra \eqref{eqn:ChargeComm} can be identified to be a free 3-step nilpotent Lie algebra known as an Engel algebra it can also be shown to be isomorphic to $A_{4,1}$ algebra of the list on Mubarakzyanov's classification of low-dimensional real Lie algebras \cite{muba, low}. The Engel algebra admits a local frame composed of two generators ${X,Y}$ whose Lie brackets commute to give
\begin{equation}
[X, Y]=Z, \quad [X, Z]=W.
\end{equation}
A natural question would be to ask how does this fermionic global symmetry arise? One can examine that the structure of $J_{3,1}$ provides a nontrivial sequence of filtrations on the tangent bundle. There is a chain of generalized eigenspaces
\begin{equation}
0 \subset \operatorname{im}\left(J_{3,1}^2\right) \subset \operatorname{im}\left(J_{3,1}\right) \subset \operatorname{ker}\left(J_{3,1}\right),
\end{equation}
where $\operatorname{im}\left(J_{3,1}^2\right)$ is one-dimensional, $\operatorname{im}\left(J_{3,1}\right)$ is two-dimensional and $\operatorname{ker}\left(J_{3,1}\right)$ is two-dimensional. Unlike the other Jordan structures, $J_{3,1}$ does not define a distribution that one could then integrate to give a foliation on the manifold instead we have a filtered manifold. We will call a Jordan structure that induces a nontrivial filtration (as opposed to a trivial one), an \textit{exotic} Jordan structure.

Recall that a filtered manifold \cite{montgomery2002tour} is defined as a manifold where the tangent bundle comes equipped with a nested sequence of filtration subspaces i.e. a filtration flag. Unlike a foliated manifold where the tangent space is still an abelian group with respect to dilations of the tangent vectors, a filtered manifold has an associated graded Lie algebra known as the symbol algebra which one can think as of nonabelian tangent space. The symbol algebra is constructed as a direct sum of elements in the flag via
\begin{equation}
\operatorname{gr}\left(T_p M\right)=\bigoplus_i F_p^i / F_p^{i-1},
\end{equation}
where $F_p^i$ denotes the terms of the filtration of the tangent space at a point p. 

For the explicit case of $J_{3,1}$, we have the filtration flag constructed by 
\begin{equation}
F_p^1=\operatorname{ker}(J), \quad F_p^2=\operatorname{ker}\left(J^2\right), \quad F_p^3=\operatorname{ker}\left(J^3\right)=T_p M .
\end{equation}
Through the rank-nullity theorem, one can check that
\begin{equation}
\operatorname{dim} F_p^1=2, \quad \operatorname{dim} F_p^2=3, \quad \operatorname{dim} F_p^3=4,
\end{equation}
which then defines the Lie-algebras
\begin{equation}
\mathfrak{g}_1=F_p^1, \quad \mathfrak{g}_2=F_p^2 / F_p^1, \quad \mathfrak{g}_3=F_p^3 / F_p^2.
\end{equation}
This Lie-algebra $\mathfrak{g}$ decomposes into a $2 \oplus 1 \oplus 1$-dimensional graded vector space of the form $\mathfrak{g}=\mathfrak{g}_1 \oplus \mathfrak{g}_2 \oplus \mathfrak{g}_3$. The associated graded lie-algebra for the Jordan structure $J_{3,1}$ is then precisely isomorphic to the Engel algebra.  In summary, the Jordan structure induces a graded Lie-algebra constructed from the filtration on the tangent bundle which then gives us an additional fermionic symmetry group on the space of fields. Interestingly, the dilations on this space acts in an anisotropic way. If the generating set of vector fields ${X,Y}$ scale by a factor $\lambda$ under dilations, then their commutator will scale by $\lambda^2$. This shows that not only are the gauge symmetries related to tropological sigma models anisotropic but their global symmetries seem to obtain a similar anisotropic scaling behavior.

\section{Nil-Equivariant Tropological Sigma Models}
\label{sec:NilTrop}
The $J_{3,1}$ model constructed in \S3.3 exhibits an additional nilpotent global symmetry that is absent in the direct $J_{2,2}$ sector. We now use this structure to construct a nil-equivariant extension of the tropological sigma model. Since the symmetry generated in \S3.3 is noncompact, the first step is to regularize it appropriately; this is achieved by quotienting by a lattice so that the relevant symmetry data descend to a compact nilmanifold. With this regularization in place, one can then formulate the corresponding equivariant BRST complex, the associated observables, and the modified selection rules.

\subsection{Compact Nilmanifold Quotients of the Engel Algebra}
In order to investigate the consequences of this additional fermionic global symmetry, we would like to turn on a background gauge field associated to the Engel algebra and what modifications the standard Gromov-Witten invariants undergo. This may be implemented by directly constructing an equivariant tropological sigma model which is equivariant with respect to the Engel algebra and so we will explicitly try to construct the Killing vector fields associated to the Engel algebra. There is one caveat however; the symmetry generators will generate a nilpotent and noncompact Lie group $G$; we will need to see how to quotient or in other words properly gauge fix this in the path integral. In order to do so, we must either find an appropriate compact subgroup of $G$. We will see that no such compact subgroups exist and we will be forced to quotient out by a lattice $\Gamma$ to achieve a proper regularization for the symmetry group.

Suppose that one chooses a compact subalgebra $K$ of $G$. We would like its Lie algebra $\mathfrak{k} \subset \mathfrak{g}$ to also be nilpotent since the Engel algebra itself is nilpotent.  As usual, we can equip the Lie algebra of any compact subgroup $K$ with an adjoint-invariant positive definite inner product and require that the subalgebra is skew adjoint with respect to the inner product. An immediate consequence of this is that any matrix representation of the nilpotent Lie algebra will have automatically be zero; this is due to the fact that any skew adjoint linear map is normal and by the spectral theorem we can unitarily diagonalize this map with a purely imaginary spectrum but nilpotency forces all eigenvalues to vanish. Consequently; the Lie subalgebra may only lie in the center of $\mathfrak{g}$ but since the group is simply connected, this means that we will only recover trivial subalgebras. The corresponding subgroups would therefore only be discrete subgroups of $G$ i.e., a lattice $\Gamma \subset G$. It is  easier to see this explicitly using a 2x2 matrix and forcing nilpotency. A quick example would be
\begin{equation}
\left(\begin{array}{cc}
0 & -\lambda \\
\lambda & 0
\end{array}\right),
\end{equation}
imposing nilpotency forces all eigenvalues to be zero.

As a result, we will instead choose to quotient out by a lattice $\Gamma$ in G. Doing so will result in a compact nilmanifold  $G / \Gamma$.  In \cite{ai1951class} has shown that there any simply connected nilpotents Lie group $G$ will admit a lattice iff its algebra admits a basis with rational structure constants. In particular, we can explicitly show that this is the case for the 4-dimensional Engel algebra via the use of canonical coordinates induced by the exponential map. Let's denote the Engel Lie algebra to be abstractly defined by generators $X_1,X_2,X_3,X_4$ that satisfy
\begin{equation}
\label{eqn:Commutators}
\left[X_1, X_2\right]=X_3, \quad\left[X_1, X_3\right]=X_4,
\end{equation}
then its associated Lie Group admits exponential coordinates $(x,y,z,w)$ of the second kind  \cite{fischer2016quantization} via
\begin{equation}
g(x, y, z, w)=\exp \left(x X_1\right) \exp \left(y X_2\right) \exp \left(z X_3\right) \exp \left(w X_4\right),
\end{equation}
one would expect that a canonical lattice would simply be given by
\begin{equation}
\Gamma_0:=\exp \left(\mathbb{Z} X_1\right) \exp \left(\mathbb{Z} X_2\right) \exp \left(\mathbb{Z} X_3\right) \exp \left(\mathbb{Z} X_4\right),
\end{equation}
however this lattice is not closed under the group law
\begin{equation}
(x, y, z, w) \cdot\left(x^{\prime}, y^{\prime}, z^{\prime}, w^{\prime}\right)=\left(x+x^{\prime}, y+y^{\prime}, z+z^{\prime}+x y^{\prime}, w+w^{\prime}+x z^{\prime}+\frac{1}{2} x^2 y^{\prime}\right).
\end{equation}
One can trace back the problem back to the $\frac{1}{2}$ that appears in the w coordinate due to the Baker-Campbell-Hausdorff formula. In order to fix this, we can either choose to rescale our basis vectors or impose a parity constraint on one of the integer coordinates; we will do the former. We begin by rescaling the 4th basis vector $X_4$ by the factor 2 and defining the basis
\begin{equation}
E_1:=X_1, \quad E_2:=X_2, \quad E_3:=X_3, \quad E_4:=2 X_4 ,
\end{equation}
we can then once again write any element of the nilpotent Engel group through the exponential map using coordinates $(x,y,z,u)$ as
\begin{equation}
g(x, y, z, u)=\exp \left(x E_1\right) \exp \left(y E_2\right) \exp \left(z E_3\right) \exp \left(u E_4\right), \quad(x, y, z, u) \in \mathbb{R}^4,
\end{equation}
the induced group law is now
\begin{equation}
\begin{aligned}
& (x, y, z, u) \cdot\left(x^{\prime}, y^{\prime}, z^{\prime}, u^{\prime}\right) \\
& =\left(x+x^{\prime}, y+y^{\prime}, z+z^{\prime}+x y^{\prime}, u+u^{\prime}+\frac{1}{2} x z^{\prime}+\frac{1}{4} x^2 y^{\prime}\right) ,
\end{aligned}
\end{equation}
the corresponding lattice is then
\begin{equation}
\Gamma=\left\{\left.\left(m, n, p, \frac{1}{4} q\right) \right\rvert\, m, n, p, q \in \mathbb{Z}\right\}=\exp \left(\mathbb{Z} E_1\right) \exp \left(\mathbb{Z} E_2\right) \exp \left(\mathbb{Z} E_3\right) \exp \left(\frac{1}{4} \mathbb{Z} E_4\right).
\end{equation}
The lattice $\Gamma$ is a rank-4 free abelian subgroup of $G$ and it induces a corresponding lattice in the Lie-algebra defined up to a rational automorphism. The associated quotient is then a compact nilmanifold known as an Engel nilmanifold $\mathcal{G}=G / \Gamma \cong \mathbb{R}^4 / \sim$ , explicitly the identification can be seen by multiplying a group element by a lattice element and it produces
\begin{equation}
(x, y, z, u) \sim\left(x+m, y+n, z+p+x n, u+\frac{1}{4} q+\frac{1}{2} x p+\frac{1}{4} x^2 n\right),
\end{equation}
for integers $n,m,p,q$. One can write these down these identifications in terms of linear integer combinations of the following generators
\begin{equation}
\begin{aligned}
& (x, y, z, u) \sim(x+1, y, z, u) \\
& (x, y, z, u) \sim\left(x, y+1, z+x, u+\frac{1}{4} x^2\right) \\
& (x, y, z, u) \sim\left(x, y, z+1, u+\frac{1}{2} x\right) \\
& (x, y, z, u) \sim\left(x, y, z, u+\frac{1}{4}\right).
\end{aligned}
\end{equation}
In order to explicitly construct the Killing vector fields, we want to equip our global symmetry group $\mathcal{G}$ with a left invariant Riemannian metric tensor. For the purposes of a background field, we may pick any left invariant metric tensor to implement this however we note in passing that if we are later interested in making the field dynamical, at first glance, it would seem impossible to gauge the Engel algebra due to the fact that any kinetic term would require an Engel adjoint invariant trace but no such adjoint invariant form exists due to the nilpotency of the Engel algebra.

We begin by computing the left-invariant Maurer-Cartan forms and we obtain
\begin{equation}
\begin{aligned}
& \theta^1=d x, \\
& \theta^2=d y ,\\
& \theta^3=d z-x d y, \\
& \theta^4=d u-\frac{1}{2} x d z+\frac{1}{4} x^2 d y,
\end{aligned}
\end{equation}
their corresponding structure equations can be seen to be $d \theta^1=0, \quad d \theta^2=0, \quad d \theta^3=-\theta^1 \wedge \theta^2, \quad d \theta^4=-\frac{1}{2}\theta^1 \wedge \theta^3$  and we can construct the left invariant Killing vector fields by solving the duality equation $\theta^a\left(K_b\right)=\delta_b^a$ which yields
\begin{equation}
    \begin{aligned}
    \label{eqn:KillingVect}
        K_1&=\partial_x, \\
        K_2&=\partial_y+x\partial_z+\frac{1}{4}x^2\partial_u, \\
        K_3&=\partial_z+\frac{1}{2}x\partial_u, \\
        K_4&=\partial_u
    \end{aligned}
\end{equation}
and they satisfy \eqref{eqn:Commutators}, with $X_4$ scaled by $1/2$. Since this frame is left-invariant, we can guarantee that this automatically descends to the nilmanifold quotient. We will use these Killing vector fields with the understanding that 
in the next subsection in order to construct an equivariant tropical topological (tropological) sigma model on $\mathbb{TP}^2$ that is equivariant with respect to the nilmanifold $\mathcal{G}$.  We will call the corresponding tropological sigma model, a \textit{Nil-Equivariant Tropological Sigma Model}.

\subsection{Nil-Equivariant Tropological Sigma Model on $\mathbb{TP}^2$}
In \cite{ewtsm}, it was shown that one can couple a relativistic topological sigma model to a background gauge field by taking the BRST operator and making it equivariant with respect to the symmetry group. In order to do this, we modify the transformation of the ghosts fields by shifting it by the Killing vector field. In doing so, we introduce the set of Nil ghosts $\phi^a$ of Grassmann degree 2. 

The field content is summarized in \tabref{tab:FieldContent}.
\begin{table}[ht]
    \centering
    \[
    \begin{array}{|c|c|c|c|}
        \hline
        \text{Field Multiplet} &
        \text{Grassmann Parity} &
        \text{Ghost Number} &
        \text{Geometric Origin} \\ \hline
        X^{i}\; =( X^{1},\Theta^{1},X^{2},\Theta^{2}) &
        \text{even} &
        0 &
        \text{Target Space Coordinates} \\ \hline
        \Psi^{i}\;= ( \psi^{1},\varphi^{1},\psi^{2},\varphi^{2}) &
        \text{odd} &
        +1 &
        \text{Foliation-Preserving Diff Ghosts} \\ \hline
        \chi^{\alpha}_{\; i}\; =(  \Xi_{1},\chi_{1},\Xi_{2},\chi_{2}) &
        \text{odd} &
        -1 &
        \text{Antighost Fields} \\ \hline
        B^{\alpha}_{\; i}\; =(  B_{1},\beta_{1},B_{2},\beta_{2}) &
        \text{even} &
        0 &
        \text{Auxiliary Fields} \\ \hline
        \phi^{a} &
        \text{even} &
        +2 &
        \text{Nil Ghosts} \\ \hline
    \end{array}
    \]
    \caption{Field content}
    \label{tab:FieldContent}
\end{table}
In order to construct a Nil equivariant tropological sigma model, we have to reintroduce the auxiliary field and lift the step 3 Engel symmetry to an off-shell symmetry of \eqref{eqn:ActionJ31}. 
Our corresponding BRST differential gets promoted to a Nil equivariant BRST differential by explicitly writing
\begin{equation}
\begin{aligned}
& Q_{\mathcal{G}}X^i=\Psi^i, \\
& Q_{\mathcal{G}} \Psi^i=\phi^a K_a^{\;i}(X), \\
& Q_{\mathcal{G}} \phi^a=0. \\
\end{aligned}
\end{equation}
One can check that this equivariant BRST differential squares to a Lie derivative along the Killing vector field $Q^2=\mathcal{L}_{\phi^a K_a}$ provided that we use the following modification of the antighost multiplet
\begin{equation}
\begin{aligned}
Q_\mathcal{G}\chi_{\;\;i}^\alpha & =\CB_{\;\;i}^\alpha, \\
Q _{\mathcal{G}}\CB_{\;\;i}^\alpha&=\mathcal{L}_{\phi^a K_a} (\chi_{\;\;i}^\alpha)=\phi^a K_a^{\;j} \partial_j \chi_{\;\;i}^\alpha+\partial_i\left(\phi^a K_a^{\;j}\right) \chi_{\;\;j}^\alpha .
\end{aligned}
\end{equation}
We recall that not all the auxiliary fields and the antighosts are zero in the gauge choice we described in \eqref{eqn:GaugeB} and \eqref{eqn:GaugeChi}. In fact the nonzero components were listed in the adapted system of coordinates in \eqref{eqn:NonZeroComp}.

The gauge fixing fermion remains as in \eqref{eqn:ActionJ31}
\begin{align}
    \psi_{\text{gf}}=\chi_{\;\;i}^\alpha E_{\;\alpha}^i-\frac{1}{2}\left(\Xi_1 B_1+\Xi_2 B_2\right)
\end{align}
and a BRST representative action for the Nil equivariant tropological sigma model is then the original action $S_0=\int\{Q,\psi_{\text{gf}}\}$,  \eqref{eqn:ActionJ31}, plus a coupling $S_{\text{Nil}}$ to the background Nil killing vector field of the form
\begin{equation}
S_{\text {Nil}}=\frac{1}{2}\int_{\Sigma} \left(\Xi_1 \mathcal{L}_{\phi^a K_a} \Xi_1+ \Xi_2 \mathcal{L}_{\phi^a K_a} \Xi_2\right),
\end{equation}
which one has to add to make sure that the full action is $Q_\CG$ invariant. 
The final localization locus is now the original tropical localization equations intersected with the fixed points of the Nil Killing vector field $K_a$, the resulting path integrals should produce Nil equivariant Gromov Witten invariants and more generally filtered Gromov Witten invariants.
\subsection{Nil-Equivariant Observables and Selection Rules}
Unlike the standard tropological sigma model, physical observables must now lie in the Nil equivariant cohomology $H^{\bullet}\left(\mathbb{T} P^2,\mathcal{G}\right)$. We recall that in the topological A model, the observables are in correspondence with BRST cohomology classes given by the pullback of deRham cohomology classes of the target. In the case for $\mathbb{CP}^1$, the quantum cohomology ring is generated by the identity operator $\mathcal{O}_{\text{Id}}$ and the symplectic form $\mathcal{O}_{\omega}$. One may explicitly construct these via the Stora-Zumino topological descent equations. On the A-model, the descent equations on $\mathcal{O}_\omega$ results in a hierarchy of observables on the worldsheet $\Sigma$, 
\begin{equation}
\begin{aligned}
d \mathcal{O}_\omega^{(0)} & =Q\mathcal{O}^{(1)}, \\
d \mathcal{O}_\omega^{(1)} & =Q \mathcal{O}^{(2)}, \\
d \mathcal{O}_\omega^{(2)} & =0 ,
\end{aligned}
\end{equation}
where $d$ is the worldsheet exterior derivative.

If one couples the A-model to topological gravity, then additional gravitational descendants $\tau_k\left(\mathcal{O}_a\right) \equiv \eta^k \wedge \mathcal{O}_a$, where $k$ is a positive integer and $a={\{\text{id},\omega\}}$, may appear at higher genus $g$ through taking additional wedge products with diffeomorphism ghosts $\eta$. As a result, every correlator in the A-model has the well-known schematic \cite{ewtg} form
\begin{equation}
\left\langle\prod_{i=1}^{n_0} \tau_{m_i}(\mathcal{O}_{\text{id}}) \prod_{j=1}^{n_1} \tau_{k_j}(\mathcal{O_\omega})\right\rangle_{g, d},
\end{equation}
the above correlator is the Gromov-Witten invariant at genus $g$ and is geometrically an intersection number on the virtual fundamental class of the moduli space of pseudoholomorphic curves of degree d $\overline{\mathcal{M}}_{g, n}(\mathbb{CP}^1, d)$. As one would expect, not all observables will be directly relevant since there is a ghost number anomaly that arises which manifests itself as an index theorem i.e. the Riemann Roch theorem of the Dolbeaut operator. Physically, this is the statement that the ghost and antighost zero modes must match and can be stated as
\begin{equation}
\operatorname{dim} \operatorname{ker} \bar{\partial}_\Phi^{\dagger}-\operatorname{dim} \operatorname{ker} \bar{\partial}_\Phi=d_X(1-g)+\int_{\Sigma} \Phi^* c_1\left(T^{1,0} \mathbb{CP}^1\right),
\end{equation}
in the case of $\mathbb{CP}^1$, this explicitly evaluates to $n_\psi-n_\chi=(1-g)+2 d$. As explained in \cite{ewtsm}, correlators that do not have the correct number of ghost and antighost insertions will automatically vanish effectively giving us a selection rule.  In the non-equivariant $\mathbb{CP}^2$ case, the Riemann Roch index theorem states $n_\psi-n_\chi=2(1-g)+3 d$.  It was shown in \cite{trsm} that the same arguments hold for the tropological sigma model and the correlators match precisely although in this case, the relevant differential operators should be deformed to their tropical counterpart. For the purposes of the Nil-equivariant tropological sigma model, we want to first begin with the family of observables that are generated by $\mathbb{CP}^2$ and then take into the account that we have an additional equivariant symmetry. 

Unlike $\mathbb{CP}^1$, the additional complex dimension allows our cohomology ring to admit one more class generated by the symplectic 2-form of $\mathbb{CP}^2$ given by the Fubini-Study metric. If we choose our degree 2 generator such that
\begin{equation}
\int_{\mathrm{CP}^1} \omega_{\mathrm{FS}}=1,
\end{equation}
where the $\mathbb{CP}^1$ is any projective line within $\mathbb{CP}^2$ then we will have an additional class given by
\begin{equation}
\int_{\mathrm{CP}^2} \omega_{\mathrm{FS}} \wedge \omega_{\mathrm{FS}}=1 .
\end{equation}
If we denote the cohomology class of the symplectic form by $h:=\left[\omega_{\mathrm{FS}}\right] \in H^2\left(\mathbb{C P}^2 ; \mathbb{Z}\right)$, then it becomes clear that the cohomology ring can be expressed as
\begin{equation}
H^{\bullet}\left(\mathbb{C P}^2 ; \mathbb{Z}\right)=\mathbb{Z}[h] /\left(h^3\right),
\end{equation}
the corresponding A-model observables can then be constructed by formally replacing the differential forms with the appropriate power of ghost fields and yields the collection $\mathcal{O}_{\text{id}}, \mathcal{O}_\omega, \mathcal{O}_{\omega \wedge \omega}$.  As previously mentioned, coupling this to topological gravity gives additional gravitational descendants that give us a general genus-g, degree-d correlator of the form
\begin{equation}
\left\langle\prod_a \tau_{m_a}\left(\mathcal{O}_{\text{id}}\right) \prod_b \tau_{n_b}\left(\mathcal{O}_\omega\right) \prod_c \tau_{p_c}\left(\mathcal{O}_{\omega \wedge \omega} \right)\right\rangle_{g, d},
\end{equation}
the Riemann-Roch theorem then gives the selection rule
\begin{equation}
\sum_a m_a+\sum_b\left(n_b+1\right)+\sum_c\left(p_c+2\right)=2(1-g)+3 d .
\end{equation}
If we turn off topological gravity, all the correlators collapse to the usual three point functions. The explicit form of these observables can once again be obtained via the Stora-Zumino topological descent equations. 

We will now compute the topological descent observables for $\mathbb{CP}^2$ since we ultimately interested in the explicit form of their equivariant analogs. The identity operator gives rise to a ghost degree zero worldsheet 0-form $\mathcal{O}_1^{(0)}=1$ which has a trivial topological descent. We will denote the deRham degree by the parenthesized superscript. We move onto the ghost degree two worldsheet 0-form for a closed $\omega$ 
\begin{equation}
\mathcal{O}_\omega^{(0)}=  \frac{1}{2}\iota_{\psi} \iota_{\psi} (X^*\omega)= \frac{1}{2}\omega_{i j} \psi^i \psi^j,
\end{equation}
where the components $\omega_{ij}$ are given by the pullback of the target symplectic form by $X$, the first step in the descent prescription then gives the ghost degree two 1-form 
\begin{equation}
\mathcal{O}_\omega^{(1)}= \frac{1}{2}\iota_{\Psi} (X^* \omega)=\omega_{ij}\psi^i\partial_\alpha X^j d\sigma^{\alpha},
\end{equation}
\begin{equation}
\mathcal{O}_\omega^{(2)}=  \frac{1}{2}X^* \omega=  \frac{1}{2}\omega_{ij}dX^i\wedge dX^j.
\end{equation}
Unlike $\mathbb{CP}^1$, we have a new seed observable that arises from the four-form $\Omega=\omega \wedge \omega$, the topological descent equations give
\begin{equation}
\mathcal{O}_{\omega\wedge\omega}^{(0)}=\frac{1}{4!}\left(\iota_{\Psi}\right)^4 (X^* \Omega)=\frac{1}{4} \omega_{i j} \omega_{k l} \psi^i \psi^j \psi^k \psi^l,
\end{equation}
\begin{equation}
\mathcal{O}_{\omega\wedge\omega}^{(1)}= \omega_{i j} \omega_{k l} d X^i \psi^j \psi^k \psi^l,
\end{equation}
\begin{equation}
\mathcal{O}_{\omega\wedge\omega}^{(2)}= \frac{3}{2} \omega_{i j} \omega_{k l} dX^i\wedge dX^j \psi^k \psi^l.
\end{equation}
In \cite{trsm}, the explicit tropical limit of these observables is able to be extracted through the Maslov dequantization and results in the components of the symplectic form being replaced by a Dirac distribution. As a result, the correlators of these observables in the tropical limit end up having the same deRham structure and ghost structure and the only thing that was needed to be demonstrated in order to completely classify the ring of observables was the cohomology of the target space. Unlike the standard approach to tropical geometry which are based on half-dimensional real algebraic geometries; the Maslov dequantization allows us to probe into tropical geometry by retaining the same dimension for the underlying geometry however it is now equipped with an additional $U(1)$ symmetry that generates the foliations on the geometry. It was shown that this additional $U(1)$ symmetry does not give rise to any additional equivariance and so for the sake of path integral calculations, we identify that the relevant cohomology for the tropical projective space $\mathbb{TP}^2$ is still $H^{\bullet}\left(\mathbb{C P}^2 , \mathbb{Z}\right)$.

In the Nil-equivariant tropological sigma model, our physical observables will still be BRST cohomology classes, however the BRST differential is now promoted to the Nil equivariant BRST differential; the relevant cohomology ring for the tropical observables is now the equivariant cohomology ring $H^\bullet\left(\mathbb{C P}^2,\mathcal{G}\right)$ where $\mathcal{G}$ is our nilmanifold global symmetry.  To begin, we need to understand how the Nilmanifold symmetry affects the symplectic form on the target space. 

There is a canonical shift in the symplectic form through the moment map. If you have a Lie group $G$ that acts on a target space $M$ by symplectomorphisms with generators $K_a$, then if the action is Hamiltonian, there exists smooth functions known as moment maps that satisfy $d \mu_a=\iota_{K_a} \omega$, in the Cartan interpretation of equivariant cohomology, one may then shift the symplectic form into the equivariant symplectic form as
\begin{equation}
\omega_{\mathrm{G}}=\omega-\mu_a \phi^a,
\end{equation}
it can be shown that this new symplectic form is equivariantly closed. 

For the case of $\mathbb{CP}^2$, one can explicitly construct these moment maps with respect to the Nilmanifold left invariant vector fields that were constructed in \eqref{eqn:KillingVect} by solving $d \mu_a=\iota_{K_a} \omega_{\mathrm{FS}}$. As a consequence, our equivariant topological observables are now all shifted by the moment map. The ghost degree 0 observable $\mathcal{O}_{\text{id}}$ still gives zero. However unlike the non-equivariant case, the ghost degree two observable is now modified to
\begin{equation}
\mathcal{O}_{\omega,\mathcal{G}}^{(0)}= \frac{1}{2} \omega_{i j} \psi^i \psi^j-\mu_a \phi^a,
\end{equation}
notice that the entire observable has ghost degree 2. The remaining hierarchy can now be computed via the equivariant topological descent equations as
\begin{equation}
\begin{aligned}
& d O_{\omega, G}^{(0)}=Q_G O_{\omega, G}^{(1)}, \\
& d O_{\omega, G}^{(1)}=Q_G O_{\omega, G}^{(2)}, \\
& d O_{\Omega, G}^{(0)}=Q_G O_{\Omega, G}^{(1)}, \\
& d O_{\Omega, G}^{(1)}=Q_G O_{\Omega, G}^{(2)},
\end{aligned}
\end{equation}
where we have the observable $\mathcal{O}_{\Omega,\mathcal{G}}$ is constructed from $\omega_G \wedge \omega_G$ that is now of ghost number $4$ and it contains term proportional $\phi^a\phi^b$ since the step 3 nilpotent Lie algebra \eqref{eqn:ChargeComm} has two non-zero structure constants. One might expect that in addition to the observables given by the equivariant symplectic form, we also have observables that are naturally coming from the additional Nil ghost field $\phi^a$ by taking a Lie-algebra trace with any integer power $n$ of the ghost field
\begin{equation}
\mathcal{O}_{\phi}^{(0)}=\operatorname{Tr \phi^n }.
\end{equation}
Nonetheless, these observables are then expected to generate a ring of observables that are in one to one correspondence with the equivariant cohomology 
$H^{\bullet}\left(\mathbb{C P}^2, \mathcal{G}\right)=\mathbb{Z}\left[\phi^a, \widetilde{h}\right] /\left(\widetilde{h}^3\right)$, where $\tilde{h}$ is now the cohomology class of the equivariant symplectic form. One immediate consequence of the moment map is that our correlators are now polynomials in Grassmann even $\phi^a$ ghost field due to the fact that our action may be supplemented by the additional term
\begin{equation}
\int_{\Sigma}\left(\Phi^* \omega+\phi^a \mu_a\right).
\end{equation}

As a simple example of the proposed refinement, one may consider the first nontrivial genus-zero, degree-one correlator on $\mathbb{CP}^2$. In the ordinary theory this is the standard degree-one invariant
\[
\Big\langle O_{\omega\wedge\omega}(z_1)\,O_{\omega\wedge\omega}(z_2)\,O_\omega(z_3)\Big\rangle_{0,1}=1.
\]
In the Nil-equivariant theory, the moment-map shift
\[
\omega_G=\omega-\mu_a\phi^a
\]
replaces the ordinary insertions by their equivariant counterparts, so the corresponding correlator becomes a polynomial in the even ghosts $\phi^a$,
\[
I^{\mathrm{Nil}}_{0,1}
=
1-\sum_a A_a\phi^a+\sum_{a,b}B_{ab}\phi^a\phi^b-\cdots .
\]
Thus the ordinary $\mathbb{CP}^2$ invariant is recovered as the constant term, while the remaining terms encode the additional dependence introduced by the filtered $J_{3,1}$ sector. The degree-one $\mathbb{CP}^2$ example and the structure of its higher-order terms are discussed in detail in Appendix~\ref{app:nilcoeff}.

\section{Conclusion and Discussions}
\label{sec:Conclusions}

In the concrete setting of $\mathbb{CP}^2$, the present paper makes explicit several features that were not worked out in the original construction of tropological sigma models. In particular, $\mathbb{CP}^2$ furnishes the first explicit example of a four-real-dimensional target arising from a complex target space in which inequivalent tropical limits can be analyzed within a single framework. The direct limit yields the familiar $J_{2,2}$ sector, whereas a nested two-step limit produces the $J_{3,1}$ sector. This provides a concrete demonstration that higher-dimensional tropicalizations need not remain purely foliated: already in this example one finds a filtered sector, together with an associated Engel-type nilpotent symmetry on the space of fields. After regularization by passage to a nilmanifold quotient, this structure in turn leads naturally to a nil-equivariant extension of the tropological sigma model. Thus, the point is not to introduce filtered geometry or Engel algebras as new mathematical structures, but to show explicitly how they arise in a tropological sigma model with target $\mathbb{CP}^2$.

We have investigated how tropological sigma models are affected by these noncompact nilpotent symmetry groups. One might wonder if these nilpotent symmetries give rise to anomalous symmetries and the important question is whether these symmetries may be gauged. At first glance, the nilpotency of the symmetry seems to suggest that no nondegenerate (in fact, nontrivial) adjoint invariant kinetic term can be written down. However, we should not be too quick in concluding that a gauging is impossible due to the fact that the existence of a non-equivariant tropological sigma model itself bases on the fact that we no longer have nondegenerate metric tensors to write kinetic terms i.e., we can nonetheless make sense of a nondegenerate quadratic form through the use of densities which allows us to in principle compute the path integrals through localization. At the very least, we would not expect any gauge anomalies due to the fact that all nilpotent Lie groups are unimodular.

It would be interesting to see the direct evaluation of such a path integral integral through supersymmetric localization to show that the theorem of Mikhalkin that tropicalization preserves Gromov-Witten invariants holds in the equivariant case where the equivariance arises due to some filtration structure on the target space. In this sense, the invariants that are defined in this model seem to go a bit further than the standard Gromov-Witten or their equivariant analogs for standard complex geometries. We conjecture that these new filtered Gromov Witten invariants might be a possible novel invariant for filtered geometries. Unlike relativistic topological sigma models, it has been previously stated in \cite{trsm} that tropological sigma models can also be directly interpreted in odd-dimensions unlike the standard relativistic case which would provide a possible way to more concretely connect the symplectic geometry of GW invariants with the filtered geometry of contact geometries. We leave this as an open question for future work.

Additionally, the Jordan-block analysis that was done in this short note is easily extended for $\mathbb{CP}^n$ where $n>2$. Already for $n=3$, one can easily see that we have much more possible filtered algebras with much more exotic symmetries that will generically be some step nilpotent Lie-algebras which are not usually studied. We have previously mentioned that complex projective space has a unique complex structure but more generic complex manifolds have a moduli space of complex structures that are allowed.  In particular, we would expect tropicalized Calabi-Yau three folds to have an interesting moduli space of Jordan structures that generically yield trivial filtrations that don't give any additional symmetry but may give rise to enhanced global symmetries when investigated at a moduli space point that describes an exotic Jordan structure in a similar respect to how enhanced gauge symmetries arise when investigating singularities of Calabi-Yau manifolds\cite{Aspinwall:1995xy}. In a similar respect, we saw that allowing for higher dimensions allowed us to consider nested Maslov dequantizations and hence nested foliations which gave rise to anisotropic nontrivial filtration structures. We expect that in higher dimensions, one may encounter much more rich structure, for example in the case of $G_2$ manifolds and once again we want to emphasize that the question about how to connect these higher dimensional nested tropicalizations to these to higher dimensional analogs of contact geometries is a promising avenue. 

\section*{Acknowledgements}
We wish to thank Ori Ganor, Petr Ho\v{r}ava and Viola Zixin Zhao for helpful discussions. We also would like to thank JHEP's Referee for useful suggestions. This work has been supported by the Leinweber Institute for Theoretical Physics at UC Berkeley.

\subsection{Appendix: Engel Algebras and the Jordan Structure $J_{3,1}$}
As discussed in the body of the paper, it is interesting to study the tropological sigma model associated to the higher order Jordan structure $J_{3,1}$ due to the appearance of an additional symmetry. We show in this appendix that the additional global symmetry is expected to arise if we look at the inverse problem.

Suppose that we were interested in simply studying the three-step nilpotent Lie algebra known as the Engel algebra, then we can know that it is generated by two vector fields ${X_1,X_2}$ whose commutators generate the Lie-algebra 
\begin{equation}
[X_1, X_2]=X_3,\quad[X_1, X_3]=X_4.
\end{equation}
The corresponding frame is known as an Engel frame and the distribution it generates is known as an Engel distribution. One may arbitrarily construct the Engel frame in local coordinates $(x,y,z,w)$ as
\begin{equation}
\begin{aligned}
& X_1=\partial_x,\\
& X_2=\partial_y+x \partial_z+\frac{1}{2}x^2 \partial_w, \\
& X_3=\left[X_1, X_2\right]=\partial_z+x\partial_w, \\
& X_4=\left[X_1, X_3\right]=\partial_w.
\end{aligned}
\end{equation}
The frame $\mathcal{B}=\left\{X_1, X_2, X_3, X_4\right\}$ is unique up to the action of a group $\mathbb{Z}_2 \times \mathbb{Z}_2$ which is generated by the following 2 elements:

$$
\begin{aligned}
& \left\{X_1, X_2, X_3, X_4\right\} \rightarrow\left\{-X_1, X_2,-X_3, X_4\right\}, \\
& \left\{X_1, X_2, X_3, X_4\right\} \rightarrow\left\{X_1,-X_2,-X_3,-X_4\right\}.
\end{aligned}
$$
A natural map to investigate is the adjoint map $\operatorname{ad}_{X_1}: Y \longmapsto\left[X_1, Y\right]$; we can see that with respect to the ordered basis above, we get a matrix representation for the adjoint which is
\begin{equation}
\left[\operatorname{ad}_{X_1}\right]_{\mathcal{B}}=\left(\begin{array}{cccc}
0 & 0 & 0 & 0 \\
0 & 0 & 0 & 0 \\
0 & 1 & 0 & 0 \\
0 & 0 & 1 & 0
\end{array}\right),
\end{equation}
\begin{equation}
\left[\operatorname{ad}_{X_2}\right]_{\mathcal{B}}=\left(\begin{array}{cccc}
0 & 0 & 0 & 0 \\
0 & 0 & 0 & 0 \\
-1 & 0 & 0 & 0 \\
0 & 0 & 0 & 0
\end{array}\right),
\end{equation}
from which we can see that $\operatorname{ad}_{X_1}$ is nilpotent of degree $3$. A similarity transformation then shows that this is simply $J_{3,1}$ in a different basis.

\subsection{Appendix: The Degree-One $\mathbb{CP}^2$ Example in the Nil-Equivariant Theory}
\label{app:nilcoeff}
We now spell out the degree-one $\mathbb{CP}^2$ example mentioned in \S4.3. The point of this example is to exhibit, in the simplest nontrivial case, how the Nil-equivariant deformation reorganizes an ordinary tropological correlator into a polynomial in the even Nil ghosts.

The ordinary cohomology ring of $\mathbb{CP}^2$ is
\[
H^\bullet(\mathbb{CP}^2;\mathbb Z)=\mathbb Z[h]/(h^3),
\qquad
h=[\omega_{FS}],
\qquad
h^2=[\omega_{FS}\wedge\omega_{FS}],
\]
with $\omega_{FS}$ normalized so that
\[
\int_{\mathbb{CP}^1}\omega_{FS}=1,
\qquad
\int_{\mathbb{CP}^2}\omega_{FS}\wedge\omega_{FS}=1.
\]
The corresponding primary observables are $O_{\mathrm{id}}$, $O_\omega$, and $O_{\omega\wedge\omega}$. For genus $g$ and degree $d$, the selection rule reads
\[
\sum_a m_a+\sum_b(n_b+1)+\sum_c(p_c+2)=2(1-g)+3d.
\]
In the case $g=0$ and $d=1$, the right-hand side is equal to $5$, so the correlator
\[
\Big\langle
O_{\omega\wedge\omega}(z_1)\,
O_{\omega\wedge\omega}(z_2)\,
O_\omega(z_3)
\Big\rangle_{0,1}
\]
is allowed, since its total degree is $2+2+1=5$. This is the ordinary genus-zero degree-one Gromov--Witten invariant of $\mathbb{CP}^2$, and its value is
\[
\Big\langle
O_{\omega\wedge\omega}(z_1)\,
O_{\omega\wedge\omega}(z_2)\,
O_\omega(z_3)
\Big\rangle_{0,1}=1.
\]
Geometrically, this is the statement that two generic points determine a unique projective line, and a generic hyperplane meets that line once.

In the Nil-equivariant theory, the BRST differential is promoted to the Nil-equivariant differential $Q_G$, and the symplectic form is replaced by its equivariant counterpart
\[
\omega_G=\omega-\mu_a\phi^a,
\]
where the $\phi^a$ are the even Nil ghosts and the $\mu_a$ are moment maps satisfying
\[
d\mu_a=\iota_{K_a}\omega_{FS}.
\]
The degree-zero observable associated with $\omega$ is therefore shifted from $O^{(0)}_\omega$ to
\[
O^{(0)}_{\omega,G}(z_i)=O^{(0)}_\omega(z_i)-m_i,
\qquad
m_i:=\mu_a(X(z_i))\phi^a.
\]
If we write
\[
\Omega:=\omega\wedge\omega,
\]
then the corresponding equivariant four-form is
\[
\Omega_G=\omega_G\wedge\omega_G,
\]
and its degree-zero observable is
\[
O^{(0)}_{\Omega,G}(z_i)
=
O^{(0)}_\Omega(z_i)-2m_i\,O^{(0)}_\omega(z_i)+m_i^2.
\]
The Nil-equivariant analogue of the ordinary degree-one correlator is therefore
\[
I^{\mathrm{Nil}}_{0,1}
:=
\Big\langle
O^{(0)}_{\Omega,G}(z_1)\,
O^{(0)}_{\Omega,G}(z_2)\,
O^{(0)}_{\omega,G}(z_3)
\Big\rangle^{J_{3,1}}_{0,1}.
\]

Setting $\phi^a=0$ gives back the ordinary insertions, and hence
\[
I^{\mathrm{Nil}}_{0,1}\big|_{\phi=0}=1.
\]
Thus the ordinary degree-one invariant appears as the constant term of the Nil-equivariant correlator.

To describe the higher-order terms, it is convenient to introduce the abbreviations
\[
\Omega_i:=O^{(0)}_\Omega(z_i),
\qquad
\omega_i:=O^{(0)}_\omega(z_i),
\qquad
\mu_a^{(i)}:=\mu_a(X(z_i)).
\]
Then
\[
m_i=\mu_a^{(i)}\phi^a,
\]
and the three insertions are
\[
O^{(0)}_{\Omega,G}(z_1)=\Omega_1-2m_1\omega_1+m_1^2,
\qquad
O^{(0)}_{\Omega,G}(z_2)=\Omega_2-2m_2\omega_2+m_2^2,
\]
\[
O^{(0)}_{\omega,G}(z_3)=\omega_3-m_3.
\]
Since each $m_i$ is linear in the even ghosts, the full product is a polynomial in the $\phi^a$, and therefore
\[
I^{\mathrm{Nil}}_{0,1}
=
1-\sum_a A_a\phi^a+\sum_{a,b}B_{ab}\phi^a\phi^b+\cdots .
\]

The constant term is the ordinary degree-one invariant. The linear coefficient is obtained by taking one first-order correction from one of the three insertions and leaving the other two at zeroth order. Equivalently, it is the coefficient of the terms linear in the $m_i$. One finds
\[
A_a=
\Big\langle
\mu_a^{(3)}\,\Omega_1\Omega_2
+
2\mu_a^{(1)}\,\omega_1\Omega_2\omega_3
+
2\mu_a^{(2)}\,\Omega_1\omega_2\omega_3
\Big\rangle^{J_{3,1}}_{0,1}.
\]
Thus $A_a$ is a correlator with a single moment-map insertion. It measures the first response of the ordinary degree-one correlator to the Nil-equivariant deformation.

The quadratic coefficient is obtained from the terms quadratic in the $m_i$. There are two kinds of contributions. The first kind comes from the quadratic terms $m_1^2$ and $m_2^2$ already present in the equivariant observables $O^{(0)}_{\Omega,G}(z_1)$ and $O^{(0)}_{\Omega,G}(z_2)$. The second kind comes from taking two linear corrections at different marked points and multiplying them together. Collecting all such terms gives
\[
\sum_{a,b}B_{ab}\phi^a\phi^b
=
\Big\langle
m_1^2\,\Omega_2\omega_3
+
\Omega_1\,m_2^2\,\omega_3
+
4m_1m_2\,\omega_1\omega_2\omega_3
+
2m_1m_3\,\omega_1\Omega_2
+
2m_2m_3\,\Omega_1\omega_2
\Big\rangle^{J_{3,1}}_{0,1}.
\]
Equivalently, after expanding $m_i=\mu_a^{(i)}\phi^a$, the coefficient $B_{ab}$ is a correlator with two moment-map insertions. It receives contributions both from a single quadratic correction inside one equivariant observable and from pairwise products of linear corrections at different marked points.

This already makes the structure of the coefficients clear. The quantities $A_a$, $B_{ab}$, and similarly the higher-order coefficients, are not formal symbols; they are specific $J_{3,1}$ correlators with moment-map insertions determined by the expansion of the equivariant observables. In this sense, the Nil-equivariant correlator reorganizes the ordinary degree-one $\mathbb{CP}^2$ invariant into a polynomial whose coefficients record the dependence of the theory on the additional equivariant data carried by the filtered sector.

Whether a particular coefficient vanishes is a dynamical question, depending on the detailed geometry of the localization locus and on the behavior of the moment maps there. The purpose of the present example is to show concretely how the ordinary degree-one invariant is recovered as the constant term, and how the first nontrivial coefficients arise from moment-map insertions in the Nil-equivariant theory. This is the sense in which the Nil-equivariant construction furnishes a polynomial refinement of the ordinary degree-one count.

\bibliographystyle{JHEP}
\bibliography{ticfpiqm.bib}

\end{document}